\documentclass{article}

\usepackage{PRIMEarxiv}
\usepackage{amsmath,amssymb}
\usepackage{booktabs}
\usepackage{multirow}
\usepackage{xcolor}
\usepackage{array} 
\usepackage[utf8]{inputenc} 
\usepackage[T1]{fontenc}    
\usepackage{hyperref}       
\usepackage{url}            
\usepackage{booktabs}       
\usepackage{amsfonts}       
\usepackage{nicefrac}       
\usepackage{microtype}      
\usepackage{lipsum}
\usepackage{fancyhdr}       
\usepackage{graphicx}       
\graphicspath{{media/}}     
\usepackage{tabularx}

\title{P1GPT: A Multi-Agent LLM Workflow Module for Multi-Modal Financial Information Analysis
}

\author{
  Chen-Che Lu \\
  Neurowatt AI \\
  Intelligent Technology Research and Development \\
  \texttt{peter@neurowatt.ai} \\
   \And
  Yun-Cheng Chou \\
  Neurowatt AI \\
  Intelligent Technology Research and Development \\
  \texttt{oro@neurowatt.ai} \\
   \And
  Teng-Ruei Chen \\
  Neurowatt AI \\
  Intelligent Technology Research and Development \\
  \texttt{luka@neurowatt.ai} \\
}

\begin{document}
\maketitle

\begin{abstract}Recent advances in large language models (LLMs) have enabled multi-agent reasoning systems capable of collaborative decision-making. However, in financial analysis, most frameworks remain narrowly focused on either isolated single-agent predictors or loosely connected analyst ensembles, and they lack a coherent reasoning workflow that unifies diverse data modalities. We introduce \textbf{P1GPT}, a layered multi-agent LLM framework for multi-modal financial information analysis and interpretable trading decision support. Unlike prior systems that emulate trading teams through role simulation, P1GPT implements a structured reasoning pipeline that systematically fuses technical, fundamental, and news-based insights through coordinated agent communication and integration-time synthesis. Backtesting on multi-modal datasets across major U.S. equities demonstrates that P1GPT achieves superior cumulative and risk-adjusted returns, maintains low drawdowns, and provides transparent causal rationales. These findings suggest that structured reasoning workflows, rather than agent role imitation, offer a scalable path toward explainable and trustworthy financial AI systems.
\end{abstract}


\section{Introduction}

\subsection{Background and Motivation}
Financial markets are inherently complex systems, where asset prices are driven by a confluence of heterogeneous signals such as corporate fundamentals, market microstructure patterns, global macroeconomic indicators, and unstructured textual data from news and social media. Traditional quantitative models and rule-based trading systems have achieved partial success, but they often rely on restrictive assumptions and struggle to adapt to rapidly evolving market conditions. In parallel, large language models (LLMs) have recently demonstrated remarkable progress in natural language understanding and domain adaptation. Notably, domain-specific initiatives such as BloombergGPT~\cite{bloomberggpt2023} and the open-source FinGPT~\cite{feng2023fingpt} illustrate how LLMs trained on financial corpora can deliver state-of-the-art performance in tasks like sentiment classification and question answering, lowering the barrier for financial AI research. These developments suggest that LLMs could play a pivotal role as intelligent assistants in financial analysis and trading.

\subsection{Challenges in Financial Trading}
Despite promising progress, deploying LLMs in real-world trading scenarios remains challenging. First, most existing approaches are single-agent systems focused on narrow tasks such as stock prediction or sentiment extraction~\cite{canllmtrade2025}, which limits their ability to capture the collaborative, multi-role workflows observed in professional investment firms~\cite{tradingagents2024}. Second, inter-agent communication is often unstructured, typically relying on free-form natural language exchanges that can introduce ambiguity and hinder consistent reasoning~\cite{finagent2024}. Third, current LLM-driven trading systems emphasize predictive accuracy but provide limited explainability, making them difficult to integrate into regulated financial environments where transparency and risk control are essential~\cite{finllama2024}. Finally, scalability is a persistent issue, as highlighted by large-scale simulation studies such as MASS~\cite{mass2025}, which underscore the difficulty of coordinating many agents and fusing multi-modal signals in complex markets.

\subsection{Contributions of P1GPT}
This work introduces \textbf{P1GPT}, a modular multi-agent LLM workflow designed to address the above challenges in financial information analysis and trading. The framework is inspired by the organizational structure of real-world trading firms and is organized into four layers: an \textit{Input Layer} for parsing heterogeneous financial and textual data, a \textit{Planning Layer} for task decomposition and agent assignment, an \textit{Analysis Layer} composed of domain-specific Intelligent Specialized Agents (ISAs) for fundamental, technical, news, and sectoral analysis, and an \textit{Integration Reasoning Layer} that fuses cross-modal outputs into coherent insights. A final \textit{Decision Layer} transforms these insights into actionable buy/hold/sell recommendations with embedded risk assessment. By combining hierarchical multi-agent orchestration~\cite{tradingagents2024}, structured inter-agent communication~\cite{finagent2024}, and depth–breadth reasoning strategies~\cite{deot2025}, P1GPT advances beyond existing single-agent financial LLMs. Our contributions are threefold: (i) we propose the first end-to-end multi-agent LLM workflow tailored for multi-modal financial analysis, (ii) we design a structured reasoning and communication protocol to improve interpretability and reliability, and (iii) we demonstrate through backtesting that P1GPT achieves superior cumulative returns and risk-adjusted performance compared with rule-based and single-agent LLM baselines.

\section{Related Work}
\subsection{Financial LLMs and Automated Assistants}
Recent advances in domain-specific large language models (LLMs) have given rise to a series of financial LLMs and assistant systems tailored for market analysis. BloombergGPT~\cite{bloomberggpt2023} introduced a 50B-parameter model trained on extensive financial corpora, demonstrating substantial improvements across multiple financial NLP benchmarks. Similarly, FinGPT~\cite{feng2023fingpt,liu2023fingpt,wang2023fingpt} provides an open-source ecosystem, later extended with Internet-scale data pipelines and instruction-tuning benchmarks, that supports benchmark evaluation and instruction tuning for financial applications, lowering the entry barrier for research and deployment. MarketSenseAI 2.0~\cite{marketsenseai2025} integrates retrieval-augmented generation with LLM agents to process earnings reports, news streams, and macroeconomic indicators, thereby enabling holistic stock analysis. Recent empirical studies further evaluated MarketSenseAI with GPT-4 in financial decision-making tasks, suggesting its potential while highlighting the challenge of consistently outperforming traditional benchmarks~\cite{marketsenseai2024}. Beyond general analysis, FinLlama~\cite{finllama2024} leverages LLMs for financial sentiment classification to assist algorithmic trading. InvestLM~\cite{investlm2023}, built on LLaMA-65B with domain-specific instruction tuning, achieves expert-level performance in investment reasoning tasks. More recent works such as Fin-R1~\cite{finr12025} highlight reinforcement learning–enhanced financial reasoning, while Agentar-Fin-R1~\cite{agentarfinr12025} extends this direction with domain-expert knowledge, trust-oriented data governance, and advanced multi-stage training strategies. In parallel, Bridging Language Models and Financial Analysis~\cite{bridging2025} explores the integration of textual, numerical, and tabular data within LLM pipelines. Collectively, these efforts highlight the growing trend of positioning LLMs as financial assistants, yet they remain largely single-agent in nature and often lack structured multi-agent workflows or scalable decision orchestration mechanisms.

\subsection{LLMs as Autonomous Traders}
Recent research has also explored the potential of large language models as autonomous trading agents that directly generate buy, hold, or sell decisions without relying on multi-agent orchestration. For instance, Can Large Language Models Trade?~\cite{canllmtrade2025} investigates whether LLMs can act as traders in simulated financial markets, testing their ability to validate classical financial theories under market dynamics. Building on this line of work, FLAG-Trader~\cite{flagtrader2025} integrates LLM-based policy networks with gradient-based reinforcement learning, demonstrating improved sequential decision-making in trading environments. Other studies highlight the role of sentiment-driven strategies, where LLMs analyze news and social media signals to produce actionable trading recommendations, as in Sentiment Trading with Large Language Models~\cite{sentimentllm2024}. Beyond direct trading signals, QuantAgent~\cite{quantagent2024} employs a self-improving loop in which LLMs generate and refine alpha factors, showing promise for automated strategy discovery. Similarly, Evaluating LLMs in Financial Tasks~\cite{llmstrategy2024} benchmarks LLMs on strategy code generation tasks, revealing both opportunities and limitations in translating natural language prompts into executable trading logic. More recently, Analysis of LLM Agent Behavior in Experimental Asset Markets~\cite{llmbehavior2025} examined how LLM traders behave in simulated bubbles and crashes, providing insights into emergent market dynamics. Additionally, MountainLion~\cite{mountainlion2025} demonstrates a multi-modal agent system in which each LLM operates as an autonomous trader, producing interpretable and adaptive strategies in volatile market conditions. Overall, these works illustrate the feasibility of LLMs functioning as standalone trading entities, yet they often lack mechanisms for structured collaboration and comprehensive multi-modal integration.

\subsection{Multi-Agent Systems in Financial Applications}
Recent studies have increasingly explored multi-agent LLM frameworks to capture the collaborative and organizational dynamics of financial decision-making. TradingAgents~\cite{tradingagents2024} models a virtual trading firm composed of analysts, researchers, traders, risk managers, and coordinators, demonstrating how role-based collaboration can enhance explainability and risk control. Similarly, the FinAgent framework, presented as A Multimodal Foundation Agent for Financial Trading~\cite{finagent2024}, integrates tool augmentation, reflective mechanisms, and diversified agents to support multi-modal financial analysis. FinVision~\cite{finvision2024} advances this direction by orchestrating multiple agents for stock market prediction, combining technical, news, and reflective modules. Beyond application-driven designs, MASS~\cite{mass2025} introduces large-scale multi-agent simulations—scaling up to 512 agents—to optimize portfolio construction and uncover scaling laws in investor modeling. Other works such as FinCon~\cite{fincon2024} mimic hierarchical financial organizations with verbal reinforcement mechanisms to improve decision robustness, while HedgeAgents~\cite{hedgeagents2025} emphasizes risk-aware collaboration through hedging strategies. More recent proposals extend multi-agent reasoning to deeper financial contexts: an analytical stage–based multi-agent framework enhanced with expert knowledge~\cite{analyticalmas2025} improves fundamental analysis, and Automate Strategy Finding with LLM in Quant Investment~\cite{automatestrategy2025} explores risk-aware multi-agent systems for quantitative strategy discovery. From a broader theoretical perspective, the Dual Engines of Thoughts (DEoT) framework~\cite{deot2025} introduces a depth–breadth integration mechanism for open-ended reasoning, offering a foundational perspective for layered multi-agent decision workflows. Together, these works illustrate the potential of multi-agent architectures to provide more scalable, interpretable, and realistic financial analysis compared with single-agent LLM approaches.

Unlike prior systems, P1GPT couples a structured, layer-wise workflow with standardized agent reports and a dedicated integration reasoning layer, enabling reproducible, auditable decisions and scalable multi-modal fusion across market regimes.

\subsection{Comparison with Existing Frameworks}

Several representative frameworks have recently explored the intersection of LLMs and financial analysis, yet each exhibits limitations that motivate the design of P1GPT. 
MarketSenseAI focuses on retrieval-augmented generation with a primarily single-agent setup, achieving solid text and data integration but lacking multi-agent coordination. 
TradingAgents models a virtual trading team with distinct roles such as analysts and risk managers, offering high interpretability but relying on free-form natural language exchanges that may reduce efficiency. 
FinAgent emphasizes tool use and multimodal inputs, yet its design remains largely experimental without a structured layered workflow. 
MASS investigates scalability with hundreds of agents, but sacrifices interpretability and practical deployment. 
FinVision combines technical and news-driven modules in a multi-agent setting, though it lacks rigorous integration and decision layers. 

\begin{table}[h]
\centering
\small
\renewcommand{\arraystretch}{1.2}
\begin{tabularx}{\linewidth}{l|X|X|X}
\toprule
\textbf{Framework} & \textbf{Architecture} & \textbf{Explainability} & \textbf{Practicality} \\
\midrule
MarketSenseAI & Single-agent + RAG; text/data/news focused & Medium (retrieval-centered) & Research-oriented, limited coordination \\
\midrule
TradingAgents & Multi-agent, role-based (analyst, trader, risk) & High (clear roles) & Simulation focus, less deployable \\
\midrule
FinAgent & Multi-agent with tool usage & Medium & Expandable, but experimental \\
\midrule
MASS & Large-scale simulation (100+ agents) & Low & Studies scalability, weak interpretability \\
\midrule
FinVision & Multi-agent (technical + news) & Medium & Academic focus, lacks integration/decision layers \\
\midrule
\textbf{P1GPT (ours)} & Layered multi-agent workflow with structured communication & High (standardized, traceable reports) & Modular, extensible, deployment-ready \\
\bottomrule
\end{tabularx}
\caption{Comparison of representative financial LLM frameworks with P1GPT.}
\label{tab:framework-comparison}
\end{table}

Compared with these frameworks, \textbf{P1GPT} provides a layered multi-agent architecture that not only supports comprehensive multimodal data integration but also enforces structured communication and standardized reporting. This design bridges the gap between academic prototypes and deployable financial systems, positioning P1GPT as a practical foundation for intelligent trading workflows.

\section{P1GPT System Overview}

\subsection{Overall System Architecture}

\begin{figure*}[h]
    \centering
    \includegraphics[width=\linewidth]{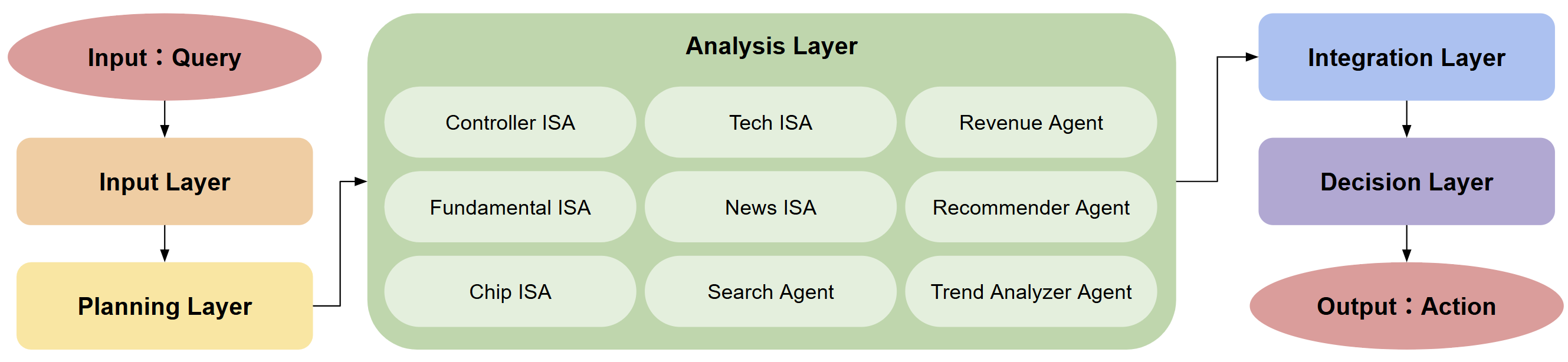}
    \caption{
    P1GPT overall workflow. Left: \textbf{Input} and \textbf{Planning} transform user queries and raw multi-modal data into an executable plan. Center: the \textbf{Analysis Layer} hosts specialized agents—Controller, Fundamental ISA, Tech ISA, Chip (Semiconductor) ISA, News ISA—and supporting modules (Search, Revenue, Trend Analyzer, Recommender). Right: the \textbf{Integration Layer} consolidates structured reports and forwards them to the \textbf{Decision Layer}, which outputs an \emph{Action} with an explanation. Arrows denote data/control flow and standardized report interfaces between layers.
    }
    \label{fig:Workflow}
\end{figure*}

P1GPT adopts a modular, layered architecture that mirrors the decision-making flow of a modern multi-agent financial analysis system. As illustrated in Figure~\ref{fig:Workflow}, the framework is organized into five distinct layers, each responsible for a critical stage of the reasoning pipeline:

\begin{enumerate}
    \item \textbf{Input Layer}:  
    Parses user queries and multi-modal financial data through entity extraction, query understanding, and data preprocessing.

    \item \textbf{Planning Layer}:  
    Decomposes tasks, assigns them to appropriate Intelligent Specialized Agents (ISA), and maps inter-agent dependencies based on query intent and data types.

    \item \textbf{Analysis Layer}:  
    Executes task-specific analysis through domain-specialized agents (e.g., Fundamental ISA, News ISA), while managing agent coordination and scheduling.

    \item \textbf{Integration Layer}:  
    Aggregates analytical outputs, performs cross-agent reasoning, and synthesizes insights into coherent intermediate summaries and structured reports.

    \item \textbf{Decision Layer}:  
    Generates actionable investment signals, conducts risk assessment and strategy planning, and produces natural language summaries as the final system output.
\end{enumerate}

This structured architecture enables modular orchestration, parallel agent execution, and complete workflow traceability. Each layer is independently extensible, ensuring that the system remains adaptable to evolving financial contexts. The detailed workflow of these layers is presented in Section 4.

\subsection{Data Sources and Collection Pipeline}

P1GPT is built to analyze multi-modal financial data, integrating a diverse set of information streams to enhance decision accuracy and robustness. The collection pipeline comprises the following major data sources:

\begin{itemize}
    \item \textbf{News and Financial Media:} Real-time news articles, press releases, and macroeconomic reports are collected from global financial news providers (e.g., Bloomberg, Reuters, Yahoo Finance) via API or RSS feeds. These sources provide timely updates on market events, company actions, and regulatory changes.
    \item \textbf{Social Media and Forums:} Posts and discussions from platforms such as X (formerly Twitter), Reddit (e.g., r/StockMarket, r/wallstreetbets), and local forums are scraped using dedicated APIs or third-party aggregators. Sentiment and engagement metrics are extracted to assess retail investor mood and crowd reactions.
    \item \textbf{Technical Market Data:} Historical and real-time price series, trading volumes, and a suite of technical indicators (e.g., MACD, RSI, SMA) are ingested from financial market data providers. This data forms the basis for technical trend analysis and quantitative modeling.
    \item \textbf{Fundamental Data:} Company financial statements, earnings reports, balance sheets, analyst estimates, and insider transactions are obtained from open databases (e.g., Yahoo Finance, Finnhub, EDGAR) and processed for fundamental valuation and risk assessment.
\end{itemize}

The pipeline automates the entire ingestion process: 
raw data is fetched at regular intervals, normalized into a unified schema, and stored in a centralized data repository. Data pre-processing includes entity recognition, timestamp alignment, duplicate removal, and language normalization.  
For textual data, natural language processing (NLP) models are used to extract relevant entities, classify sentiment, and generate summaries.  
All processed and standardized data is then made available to the relevant agents via a shared access layer, ensuring that each agent can query the latest information for its analytical tasks.

This comprehensive and automated pipeline enables P1GPT to maintain an up-to-date and holistic view of the financial environment, supporting timely and well-informed multi-agent reasoning.

\subsection{Core Modules and Agent Roles}

P1GPT employs a modular multi-agent design, inspired by the structure of real-world financial analysis teams and the architecture of the Summarizer system. Each agent in the system is responsible for a specialized analytical task, contributing to a coordinated and explainable decision-making pipeline. These agents are organized into two categories: \textit{Intelligent Specialized Agents (ISA)} and \textit{Supporting Agents}.

\subsubsection{Intelligent Specialized Agents (ISA)}
The ISA family consists of domain-specialized agents designed to handle core financial analysis functions. Each ISA processes a particular modality or market perspective using LLMs and financial tools:

\begin{itemize}
    \item \textbf{Controller ISA}: Serves as the central coordinator of the agent workflow. It assigns tasks, manages execution order, and integrates analytical outputs to ensure coherent reasoning and final synthesis.

    \item \textbf{Fundamental Analysis ISA}: Analyzes company fundamentals, including earnings reports, balance sheets, valuation ratios, and analyst expectations. It provides long-term investment insights based on corporate financial health.

    \item \textbf{Technical Analysis ISA}: Interprets market signals from historical price and volume data by applying technical indicators such as MACD, RSI, and moving averages. It supports short- to medium-term trading strategies.

    \item \textbf{Semiconductor ISA}: Focuses on the semiconductor and supply chain domain, aggregating and interpreting industry news, market trends, and innovation signals to assess sector-specific opportunities and risks.

    \item \textbf{News Intelligence ISA}: Aggregates financial news, regulatory updates, and press releases. It applies NLP techniques to summarize content, extract sentiment, and identify events that may influence market dynamics.
\end{itemize}

\subsubsection{Supporting Agents}
These agents enhance the capabilities of the core ISA modules by providing external retrieval, forecasting, and decision support functionalities:

\begin{itemize}
    \item \textbf{External Search Agent}: Performs real-time information retrieval from external knowledge bases and APIs (e.g., Perplexity AI) to supplement ISA analysis with up-to-date facts and context.

    \item \textbf{Revenue Forecasting Agent}: Utilizes statistical models and historical financial data to predict future revenue or earnings figures for target companies, aiding in valuation and scenario testing.

    \item \textbf{Investment Recommendation Agent}: Synthesizes the outputs of all analytical agents to formulate actionable investment decisions (e.g., buy/sell/hold), along with confidence scores and supporting rationales.

    \item \textbf{Market Trend Analysis Agent}: Tracks broader market indices, trading volumes, and price momentum to identify macroeconomic trends, sectoral rotations, and long-term patterns.
\end{itemize}

All agents operate as independent modules with defined input/output interfaces and communicate through a structured protocol. Their outputs are formatted into standardized reports, which are aggregated and evaluated by the Controller ISA and the decision-making layer. This architecture ensures flexibility, scalability, and robust multi-modal reasoning across complex financial environments.

\subsection{Integration Strategy and Data Fusion}

To maximize the value of multi-modal information, P1GPT implements a structured integration strategy that systematically fuses the outputs of all specialized agents. Each analytical agent produces a standardized report summarizing its findings, which are submitted to the Overall ISA Agent for aggregation and cross-comparison.

The integration process involves both rule-based and LLM-driven fusion techniques. Quantitative results from technical and fundamental analyses are normalized and combined with qualitative insights from news and sentiment agents. A centralized decision module evaluates the consistency and reliability of each agent's input, assigning dynamic weights or confidence scores according to the context and recent agent performance.

Where discrepancies or conflicts exist between agent assessments, the system may trigger additional analysis (e.g., calling the Perplexity Search Agent for external verification or requesting deeper analysis from a relevant agent). This iterative information refinement ensures that all available perspectives are considered before decision-making.

The final, integrated view supports robust and explainable trading decisions by synthesizing numerical scores, textual rationales, and trend predictions into a single actionable recommendation. This data fusion approach enhances both the breadth and depth of system reasoning, enabling the P1GPT framework to adapt flexibly to complex and rapidly changing market environments.

\section{Layered Workflow and Communication}

The P1GPT system is designed around a layered architecture that facilitates modularity, transparency, and collaboration among agents. From the initial user query to the final investment recommendation, each layer transforms, enriches, and routes information to the next stage of the pipeline. This section details the complete operational flow across the five core layers.

\subsection{Input Layer}

The Input Layer is responsible for transforming raw inputs—both user queries and external data streams—into structured representations suitable for downstream agent processing. It ensures that all subsequent reasoning is grounded in clean, interpretable, and context-aware information. Its core objectives include:

\begin{itemize}
    \item \textbf{User Query Understanding}  
    Natural language instructions are parsed to extract intent, target entities (e.g., tickers, indicators), and temporal scope using domain-adapted NLP models.

    \item \textbf{Entity and Feature Extraction}  
    Financial entities such as companies, metrics, and events are identified and tagged from both queries and raw textual data (e.g., news articles, forum posts).

    \item \textbf{Data Preprocessing and Harmonization}  
    Structured (e.g., price data), semi-structured (e.g., filings), and unstructured sources (e.g., social media) are normalized into a unified format, with noise removed and timestamps aligned.

    \item \textbf{Input Structuring for Planning Layer}  
    Parsed queries and preprocessed data are encapsulated into standardized input objects annotated with semantic metadata, enabling efficient consumption by the Planning Layer and Intelligent Specialized Agents.
\end{itemize}

By serving as a semantic gateway for both user intent and real-world market signals, the Input Layer ensures accuracy, clarity, and consistency throughout the P1GPT analytical workflow.

\subsection{Planning Layer}

The Planning Layer translates structured inputs from the Input Layer into an executable multi-agent workflow.  
It determines the analytical scope, decomposes complex tasks, and orchestrates agent collaboration to ensure comprehensive and efficient coverage of the problem space.  
Its key functions include:

\begin{itemize}
    \item \textbf{Task Decomposition}  
    Breaks down high-level analysis objectives into granular subtasks (e.g., fundamental analysis, sentiment extraction, trend detection) that can be executed independently or in sequence.

    \item \textbf{Agent Matching}  
    Maps each subtask to the most suitable Intelligent Specialized Agent (ISA) or supporting agent based on domain expertise, required data modality, and computational capabilities.

    \item \textbf{Dependency Mapping}  
    Establishes inter-task dependencies and execution order, enabling the system to run agents in parallel when possible while enforcing sequential processing where data flow dictates.

    \item \textbf{Dynamic Workflow Adaptation}  
    Adjusts task allocation in real time if new data becomes available, user intent changes, or intermediate results suggest alternative analytical paths.
\end{itemize}

Through precise decomposition, optimal agent assignment, and dependency-aware scheduling,  
the Planning Layer ensures that every query triggers a coordinated, resource-efficient multi-agent reasoning process.

\subsection{Analysis Layer}

The Analysis Layer is the operational core of P1GPT, where domain-specialized reasoning is performed by both Intelligent Specialized Agents (ISAs) and supporting agents.  
Upon receiving the execution plan from the Planning Layer, each assigned agent processes its designated data segment using domain-specific methodologies and models.

ISAs cover critical analytical domains such as fundamentals, technical indicators, sector-specific intelligence (e.g., semiconductors), and market news sentiment.  
Each agent operates as an autonomous processing unit, equipped with tailored analytical pipelines—ranging from financial ratio evaluation and time-series modeling to natural language processing for event extraction. Supporting agents, such as external search and revenue forecasting modules, are invoked when additional contextual knowledge or predictive insights are required.

Inter-agent coordination is enabled through a shared communication protocol. Agents can request clarification from peers, exchange intermediate findings, and collaboratively resolve conflicting interpretations.  
This coordination mechanism is particularly important when dealing with multi-modal inputs, where insights from different domains must be reconciled before integration.

By combining independent domain expertise with controlled information exchange, the Analysis Layer ensures that the overall reasoning process remains both comprehensive and internally consistent.  
Its outputs—structured analytical reports annotated with confidence scores and explanatory rationales—form the foundation for the subsequent integration and decision-making stages.

\subsection{Integration Layer}

The Integration Reasoning Layer consolidates the heterogeneous outputs generated by the Analysis Layer into a unified, coherent analytical perspective.  
This stage is responsible for aligning, reconciling, and synthesizing results across multiple modalities and domains to ensure that subsequent decision-making is both informed and explainable.

Integration begins with the collection of structured reports from all participating agents. These reports contain standardized elements such as key findings, supporting evidence, confidence scores, and domain-specific rationale.  
An alignment process maps the outputs to shared reference entities—such as stock tickers, timeframes, or economic indicators—allowing cross-domain comparisons and eliminating redundancies.

Reasoning within this layer is performed through a combination of deterministic rules and LLM-driven synthesis. Rule-based logic is employed to enforce domain priorities and resolve conflicts (e.g., prioritizing fundamental analysis over short-term sentiment in long-horizon strategies), while large language models are used to produce higher-level interpretations that integrate quantitative metrics with qualitative insights.

When conflicting or uncertain conclusions emerge, the layer may trigger iterative feedback loops with the Analysis Layer, requesting clarification, additional computations, or external validation. This ensures that discrepancies are addressed before final recommendations are produced.

By merging structured numerical signals with narrative reasoning, the Integration Reasoning Layer provides a complete and context-aware assessment of the financial environment, serving as the critical bridge between raw analysis and actionable decision-making.

\subsection{Action Decision Layer}

The Action Decision Layer transforms the synthesized insights from the Integration Reasoning Layer into concrete, actionable outputs for end-users or automated trading systems.  
Its primary objective is to bridge the gap between analytical assessment and operational execution while maintaining transparency and traceability.

This layer is driven by specialized decision-making agents, most notably the Investment Recommendation Agent and the Market Trend Analysis Agent.  
The former evaluates the integrated signals against predefined investment strategies, risk tolerance parameters, and market conditions to produce explicit trade recommendations—such as Buy, Sell, or Hold—accompanied by confidence scores and justifications.  
The latter focuses on identifying and validating mid- to long-term market trends, providing contextual guidance that can influence position sizing and portfolio adjustments.

Decisions are generated through a hybrid process:  
quantitative thresholds and risk metrics ensure that recommendations meet objective performance and safety criteria, while LLM-based reasoning modules articulate the rationale in natural language, enabling human users to understand and audit the underlying logic.

Where execution is automated, this layer interfaces directly with trading platforms or portfolio management systems through secure APIs, ensuring timely and accurate order placement.  
In human-in-the-loop scenarios, the recommendations are presented via an interactive dashboard, allowing analysts or traders to review, adjust, or override system-generated actions.

By tightly coupling analytical rigor with operational delivery, the Action Decision Layer ensures that P1GPT’s outputs are not only theoretically sound but also practically implementable in dynamic financial environments.

\section{Experiments}

\subsection{Simulation Setup}

To evaluate the performance of \textbf{P1GPT} in realistic financial trading scenarios, we conduct comprehensive backtesting simulations using multi-modal historical data.  
The simulation focuses on three representative U.S.\ equities—\textbf{AAPL}, \textbf{GOOGL}, and \textbf{TSLA}—which collectively span major innovation-driven sectors: consumer technology, digital platforms, and electric vehicles.  
These companies were selected because they represent distinct yet interconnected domains that are highly sensitive to macroeconomic shifts, supply chain policies, and investor sentiment.  
Analyzing these stocks allows the evaluation of P1GPT’s adaptability across heterogeneous market structures and volatility regimes.
The evaluation period spans from \textbf{February 1, 2025} to \textbf{September 30, 2025}, a phase marked by notable post-election market uncertainty and renewed trade policy dynamics.  

In particular, the reinstatement of import tariffs under the Trump administration in early 2025 triggered sector-specific volatility—especially within technology and semiconductor supply chains—providing a realistic stress test for the system’s reasoning-driven decision process.  
This window also includes several quarterly earnings releases and sentiment reversals, enabling assessment of how P1GPT responds to rapidly evolving multi-modal financial contexts.
  
For each trading day, the system receives a structured prompt that includes the current market context and a target stock ticker.  
All data are preprocessed through the collection and normalization pipeline described in Section~3.2 to ensure consistency and temporal alignment.  
The simulation strictly avoids lookahead bias—agents can only access information available prior to the decision timestamp.

\vspace{0.5em} 
\noindent 
\textbf{Baselines.} To benchmark the performance of P1GPT, we compare it against five established strategies following the TradingAgents evaluation protocol: \begin{itemize} 
    \item \textbf{Buy \& Hold (B\&H):} Passive investment throughout the simulation window. 
    \item \textbf{MACD:} Moving Average Convergence Divergence crossover strategy. 
    \item \textbf{KDJ + RSI:} Momentum-based signals combining stochastic oscillator and RSI thresholds to detect overbought/oversold conditions. 
    \item \textbf{ZMR:} Zero-Mean Reversion strategy using deviations from an equilibrium price as trading triggers. 
    \item \textbf{SMA:} Simple moving-average cross strategy (10-day vs.\ 20-day windows). 
\end{itemize}

These baselines provide diverse perspectives on market behavior—ranging from passive exposure to active momentum and reversion-based trading—allowing a fair evaluation of P1GPT’s reasoning-driven decision process.

\vspace{0.5em}
\noindent
\textbf{Evaluation Metrics.}  
Four standard financial metrics are used to evaluate the profitability, risk, and stability of each strategy:

\begin{itemize}
    \item \textbf{Cumulative Return (CR):}
    \[
        CR = \frac{V_{\text{end}} - V_{\text{start}}}{V_{\text{start}}} \times 100\%
    \]
    where $V_{\text{start}}$ and $V_{\text{end}}$ denote initial and final portfolio values, respectively.

    \item \textbf{Annualized Return (AR):}
    \[
        AR = \left(\frac{V_{\text{end}}}{V_{\text{start}}}\right)^{1/N} - 1
    \]
    where $N$ represents the number of years in the simulation period.

    \item \textbf{Sharpe Ratio (SR):}
    \[
        SR = \frac{\overline{R} - R_f}{\sigma}
    \]
    where $\overline{R}$ is the average daily return, $R_f$ the risk-free rate (approximated by 3-month U.S.\ Treasury yield), and $\sigma$ the standard deviation of daily returns.

    \item \textbf{Maximum Drawdown (MDD):}
    \[
        MDD = \max_{t \in [0, T]} \left(\frac{\text{Peak}_t - \text{Trough}_t}{\text{Peak}_t}\right) \times 100\%
    \]
    which represents the largest observed loss from a peak to a trough within the simulation window.
\end{itemize}

\subsection{Backtesting Simulation}

For each trading day $t$, P1GPT receives a structured query of the form:  
\texttt{"Given today's market conditions, should I buy, sell, or hold [TICKER]?"}  
This query is processed through the system’s hierarchical reasoning pipeline, activating specialized agents in the Fundamental, Technical, News, and Sentiment domains.  
Each agent contributes domain-specific insights that are synthesized within the Integration and Decision Layers to produce a final investment signal.

The system outputs a single trading decision—\textbf{Buy}, \textbf{Sell} or \textbf{Hold}—accompanied by a concise natural-language explanation summarizing the reasoning process.  
No numerical confidence score is produced; instead, interpretability is achieved through the transparent textual rationale and structured inter-agent reports.

A simple execution policy governs portfolio operations:
\begin{itemize}
    \item \textbf{Buy:} Enter a position at the same day’s closing price.
    \item \textbf{Sell:} Exit any currently held position.
    \item \textbf{Hold:} Maintain the existing position without modification.
\end{itemize}

The backtesting assumes no transaction costs, no leverage, and a single-stock position constraint per run.  
Daily portfolio values are aggregated to compute CR, AR, SR, and MDD across the test horizon.  
The resulting statistics enable an objective comparison between P1GPT and baseline strategies in terms of profitability, volatility control, and drawdown resilience.

\section{Results, Analysis, and Discussion}

\begin{figure*}[h]
    \centering
    \includegraphics[width=\linewidth]{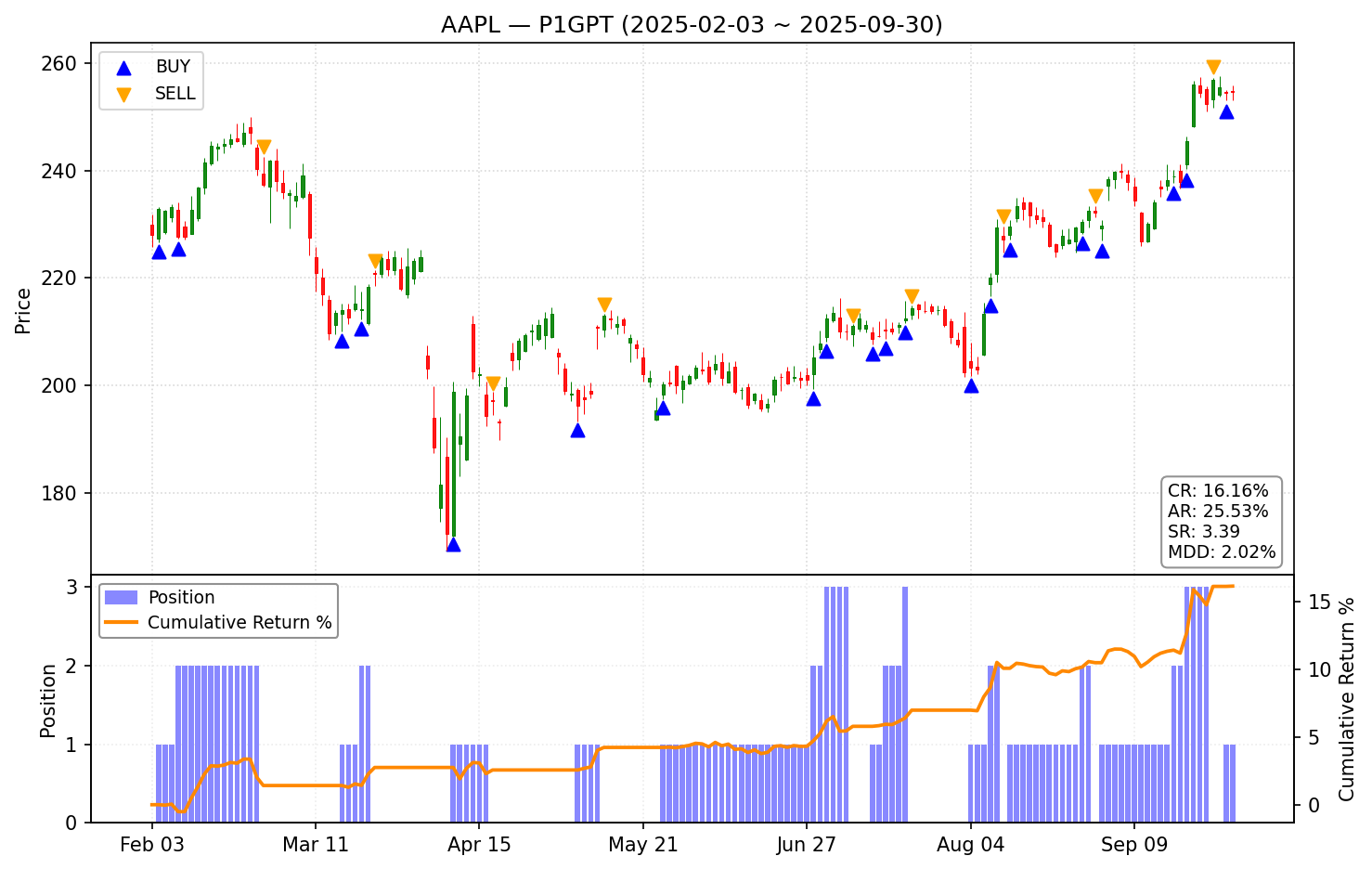}
    \caption{
    Trading trajectory of P1GPT on AAPL (Feb.~3–Sep.~30,~2025), showing model-generated buy (blue) and sell (orange) signals, along with position and cumulative return evolution.
    }
    \label{fig:aapl_perf}
\end{figure*}

\begin{figure*}[h]
    \centering
    \includegraphics[width=\linewidth]{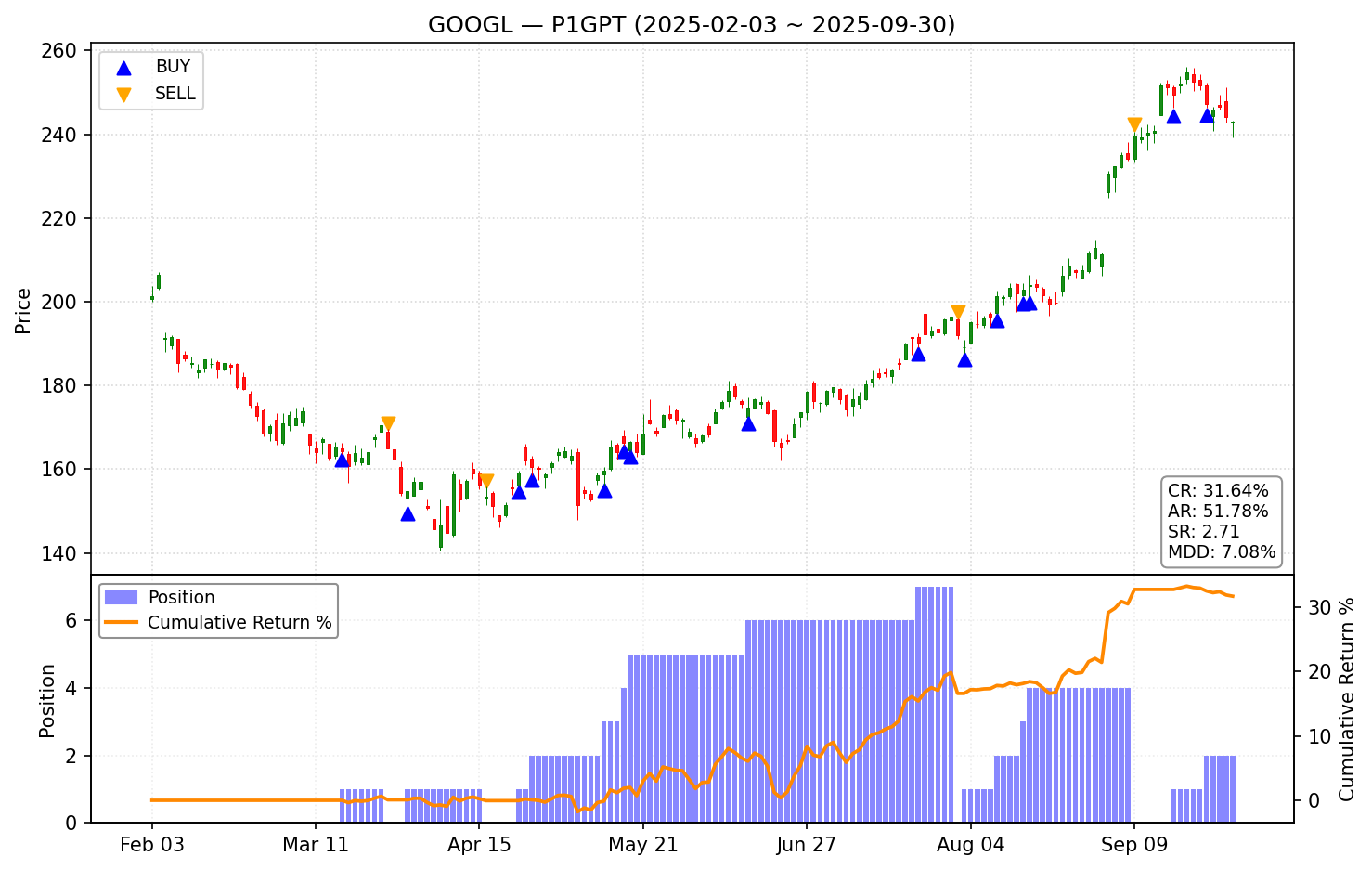}
    \caption{
    Trading trajectory of P1GPT on GOOGL (Feb.~3–Sep.~30,~2025), showing model-generated buy (blue) and sell (orange) signals, along with position and cumulative return evolution.
    }
    \label{fig:googl_perf}
\end{figure*}

\begin{figure*}[h]
    \centering
    \includegraphics[width=\linewidth]{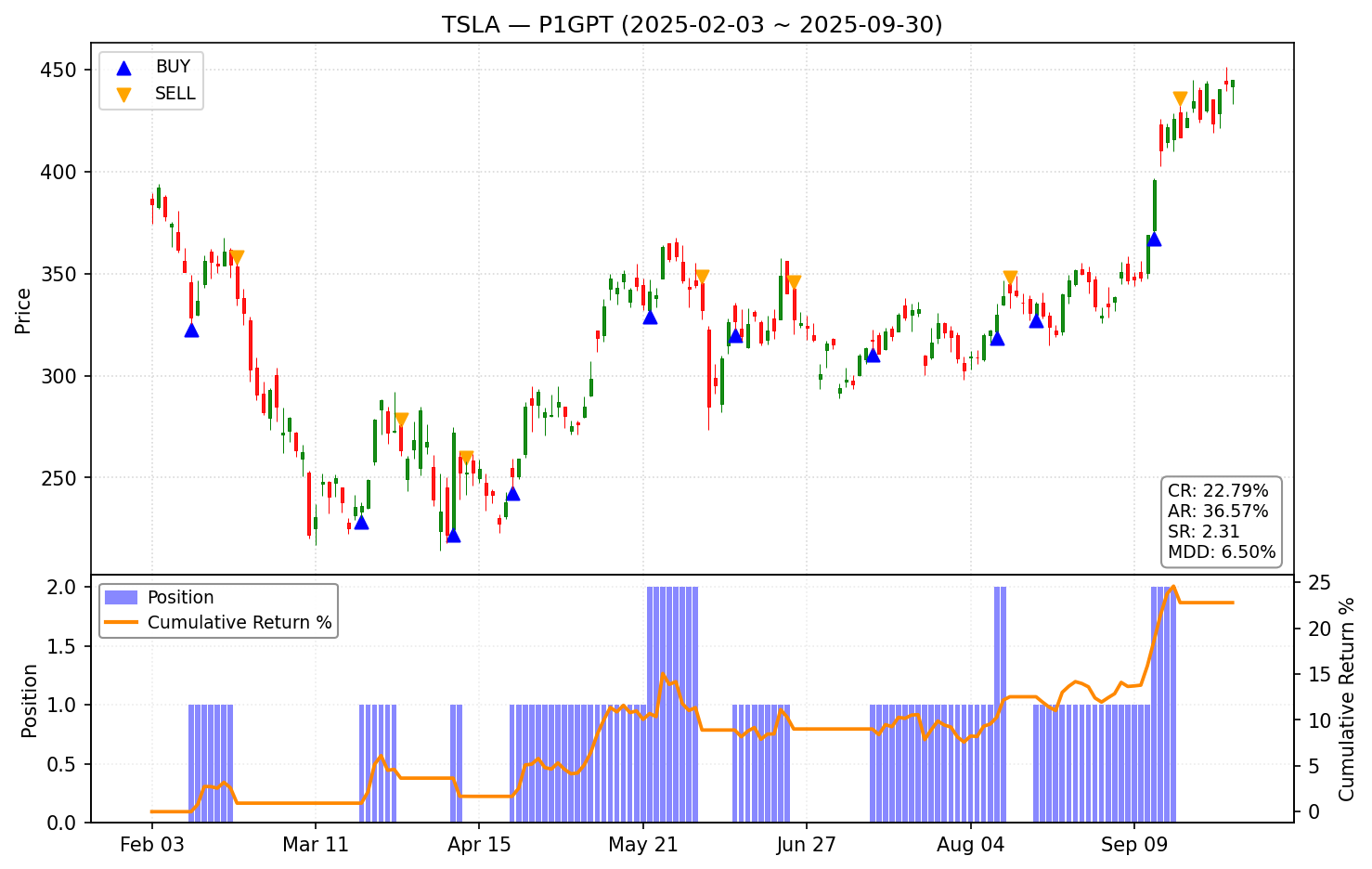}
    \caption{
    Trading trajectory of P1GPT on TSLA (Feb.~3–Sep.~30,~2025), showing model-generated buy (blue) and sell (orange) signals, along with position and cumulative return evolution.
    }
    \label{fig:tsla_perf}
\end{figure*}

\begin{figure*}[h]
    \centering
    \includegraphics[width=\linewidth]{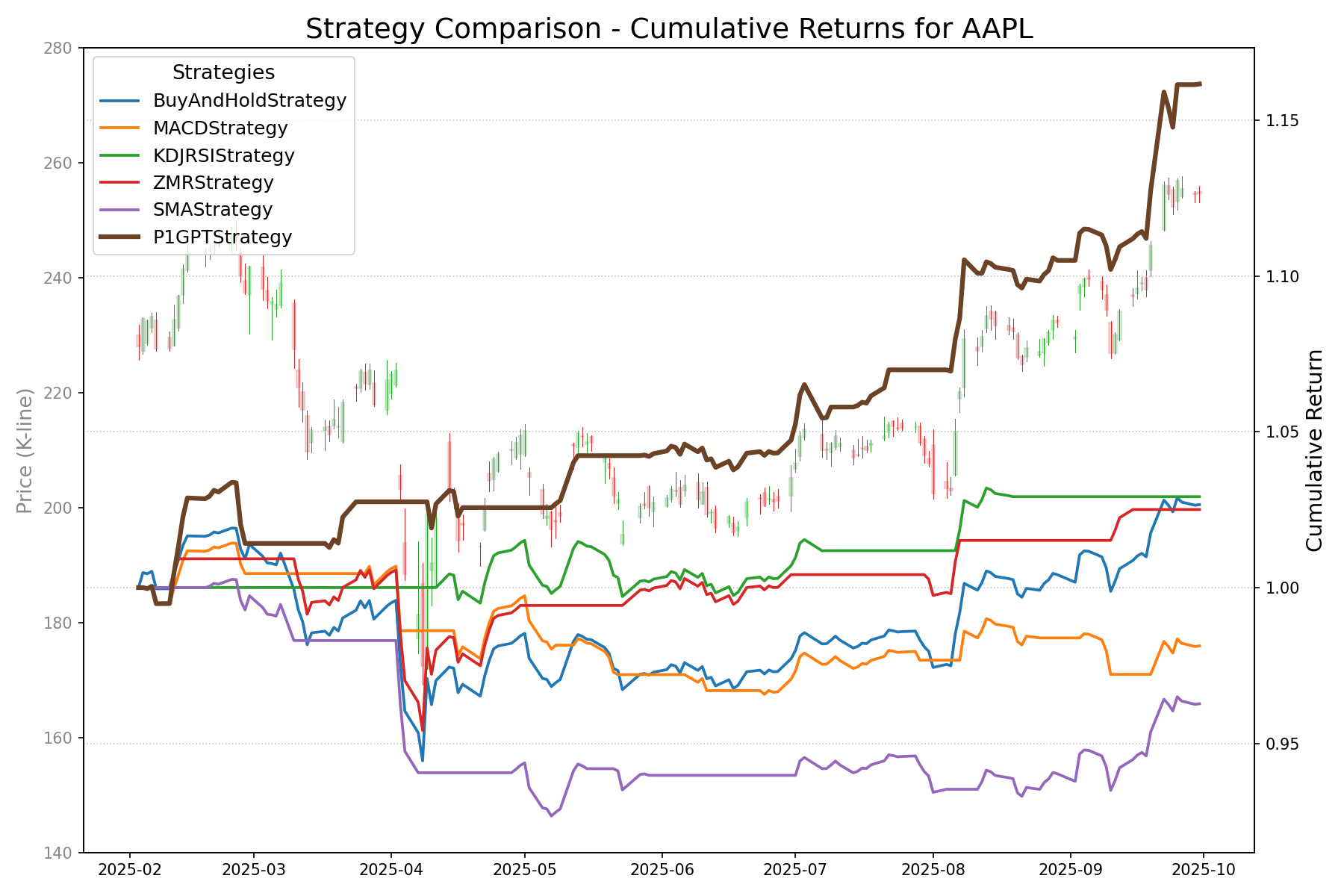}
    \caption{ Cumulative Returns on AAPL using P1GPT. The figure shows the performance comparison of our model against baseline approaches for Apple Inc. stock analysis.
    }
    \label{fig:strategy_comparison}
\end{figure*}

\subsection{Quantitative Performance Evaluation}

The quantitative outcomes are summarized in Table~\ref{tab:perf_all}, 
accompanied by Figures~\ref{fig:aapl_perf}–\ref{fig:strategy_comparison}, 
which illustrate P1GPT’s equity growth trajectories on AAPL, GOOGL, and TSLA.  
Together, they reveal that \textbf{P1GPT} maintains a well-calibrated balance between profitability and risk, 
consistently outperforming traditional rule-based strategies across diverse market environments.

\begin{table*}[ht]
\centering
\small 
\setlength{\tabcolsep}{3pt} 
\renewcommand{\arraystretch}{1.05} 
\caption{Performance comparison across all methods using four evaluation metrics. 
Results highlighted in \textcolor{green!50!black}{green} represent the best-performing statistic 
(higher is better for CR/AR/SR; lower is better for MDD).}
\label{tab:perf_all}
\begin{tabular}{llcccccccccccc}
\toprule
\multirow{2}{*}{\textbf{Categories}} & 
\multirow{2}{*}{\textbf{Models}} & 
\multicolumn{4}{c}{\textbf{AAPL}} & 
\multicolumn{4}{c}{\textbf{GOOGL}} & 
\multicolumn{4}{c}{\textbf{TSLA}} \\
\cmidrule(lr){3-6} \cmidrule(lr){7-10} \cmidrule(lr){11-14}
& & \textbf{CR\%$\uparrow$} & \textbf{AR\%$\uparrow$} & \textbf{SR$\uparrow$} & \textbf{MDD\%$\downarrow$}
  & \textbf{CR\%$\uparrow$} & \textbf{AR\%$\uparrow$} & \textbf{SR$\uparrow$} & \textbf{MDD\%$\downarrow$}
  & \textbf{CR\%$\uparrow$} & \textbf{AR\%$\uparrow$} & \textbf{SR$\uparrow$} & \textbf{MDD\%$\downarrow$} \\
\midrule
\textbf{Market}
& B\&H      & 2.66 & 4.07 & 0.56 & 7.33 & 4.19 & 6.42 & 1.05 & 6.41 & 6.10 & 9.41 & 0.52 & 16.89 \\
\midrule
\textbf{Rule-based}
& MACD      & -1.87 & -2.83 & -0.59 & 4.79 & 2.67 & 4.09 & 0.83 & 4.11 & 7.36 & 11.38 & 0.88 & 10.69 \\
& KDJ+RSI   & 2.92 & 4.46 & 1.24 & \textcolor{green!50!black}{1.78} & 0.98 & 1.49 & 0.60 & \textcolor{green!50!black}{1.28} & 8.65 & 13.43 & 1.13 & \textcolor{green!50!black}{6.37} \\
& ZMR       & 2.50 & 3.83 & 0.64 & 5.46 & -0.62 & -0.94 & -0.22 & 4.69 & 3.61 & 5.53 & 0.41 & 15.60 \\
& SMA       & -3.73 & -5.60 & -1.03 & 7.57 & 8.23 & 12.76 & \textcolor{green!50!black}{2.90} & 1.32 & 6.31 & 9.73 & 0.68 & 11.43 \\
\midrule
\textbf{Ours}
& \textbf{P1GPT} 
& \textcolor{green!50!black}{16.16} & \textcolor{green!50!black}{25.53} & \textcolor{green!50!black}{3.38} & 2.02
& \textcolor{green!50!black}{31.64} & \textcolor{green!50!black}{51.78} & 2.71 & 7.08
& \textcolor{green!50!black}{22.79} & \textcolor{green!50!black}{36.57} & \textcolor{green!50!black}{2.31} & 6.50 \\
\bottomrule
\end{tabular}
\end{table*}

\subsubsection{Cumulative and Annual Returns}
P1GPT achieves the highest profitability across all tested assets, 
with cumulative returns exceeding 16\% on AAPL, 31\% on GOOGL, and 22\% on TSLA.  
Its annualized returns surpass the best rule-based baselines by over 20 percentage points on average.  
This improvement stems from its hierarchical reasoning framework, which integrates multi-modal signals— 
including earnings trends, technical momentum, and sentiment shifts—allowing adaptive repositioning during changing market phases.  
Unlike static threshold systems such as SMA or MACD, P1GPT’s dynamic synthesis produces smoother return curves and sustained compounding over time.

\subsubsection{Sharpe Ratio}
P1GPT maintains strong risk-adjusted performance, achieving Sharpe ratios of roughly 3.4 (AAPL), 2.7 (GOOGL), and 2.3 (TSLA).  
Although SMA slightly exceeds it on GOOGL, P1GPT exhibits greater cross-asset consistency, 
reflecting a deliberate risk–return calibration rather than volatility suppression.  
By tolerating measured fluctuations when they contribute to higher expected payoff, 
the system converts information advantage into reward without relying on excessive leverage or reactive timing.

\subsubsection{Maximum Drawdown}
The framework’s maximum drawdowns remain controlled—around 2\% for AAPL and below 7\% for the other assets— 
demonstrating resilience under market stress.  
While certain conservative baselines achieve marginally lower drawdowns (e.g., KDJ+RSI on GOOGL at 1.3\%), 
they do so at the cost of diminished returns.  
P1GPT’s behavior reflects a balanced posture: accepting limited short-term losses to capture larger long-term gains, 
achieving stability through informed adaptability rather than rigid constraint.

\subsubsection{Cross-Asset Robustness}
Across the three sectors—consumer technology, digital services, and electric vehicles—P1GPT consistently outperforms baseline strategies in cumulative and annualized returns 
while maintaining Sharpe ratios above 2.0 and drawdowns within moderate bounds.  
This consistency underscores its generalization capability: 
the reasoning-driven, multi-agent architecture transfers effectively across market regimes, 
delivering interpretable and risk-aware performance rather than over-optimized results specific to one asset or indicator.

\subsection{Qualitative Behavioral Analysis}

\subsubsection{Overview of Behavioral Patterns}

Figures~\ref{fig:aapl_perf}–\ref{fig:tsla_perf} visualize the trading trajectories of P1GPT across three representative equities.  
Across all assets, the system exhibits a consistent behavioral pattern characterized by \textit{adaptive re-entry}, \textit{risk-aware exit}, and \textit{volatility-conscious exposure control}.  
Rather than performing frequent short-term trades, P1GPT demonstrates a structured rhythm of position building and unwinding that reflects reasoning-driven judgment rather than reactive signal following.

For \textbf{AAPL}, the system repeatedly re-enters positions shortly after localized drawdowns and avoids holding through extended downtrends.  
This behavior suggests a disciplined recovery strategy—reallocating exposure once technical and fundamental conditions realign—leading to steady cumulative gains with limited volatility.  
In contrast to rule-based methods that rely on fixed thresholds, P1GPT’s actions adapt dynamically to changing market regimes, as seen in its avoidance of early entries during the March downturn and its timely re-engagement around July–August when momentum stabilized.

For \textbf{GOOGL}, P1GPT’s trade sequence reveals a gradual accumulation phase during the mid-year recovery, followed by profit-taking near short-term peaks.  
The position histogram indicates that the system increased exposure as confidence signals strengthened across multiple modalities (e.g., fundamental growth, sentiment improvement), aligning with the broader uptrend.  
Although its drawdown (\textbf{7.08\%}) was higher than that of AAPL, the cumulative return (\textbf{31.64\%}) shows that the model consciously accepted controlled risk in exchange for higher opportunity capture.

On \textbf{TSLA}, the agent demonstrates cautious entry behavior during the volatile early period, favoring shorter holding intervals and early exits amid downward pressure.  
As sentiment and technical indicators reversed around mid-year, P1GPT transitioned toward longer holding durations and a pronounced upward exposure in late Q3.  
This behavioral shift illustrates its capacity for \textit{contextual reasoning}: the model interprets fundamental recovery and sentiment reversal as sufficient justification for re-entry, achieving efficient timing without overtrading.

Taken together, these behavioral patterns illustrate P1GPT’s explainable investment logic:  
it internalizes risk–reward trade-offs through cross-agent consensus, forming a decision process that combines macro context, sector-specific trends, and short-term signals.  
Such reasoning-based adaptivity explains why the model achieves consistent profitability under heterogeneous market conditions while maintaining interpretable decision transparency.

\subsubsection{Case-Level Event Analysis}

\begin{figure*}[t]
    \centering
    \includegraphics[width=0.95\linewidth]{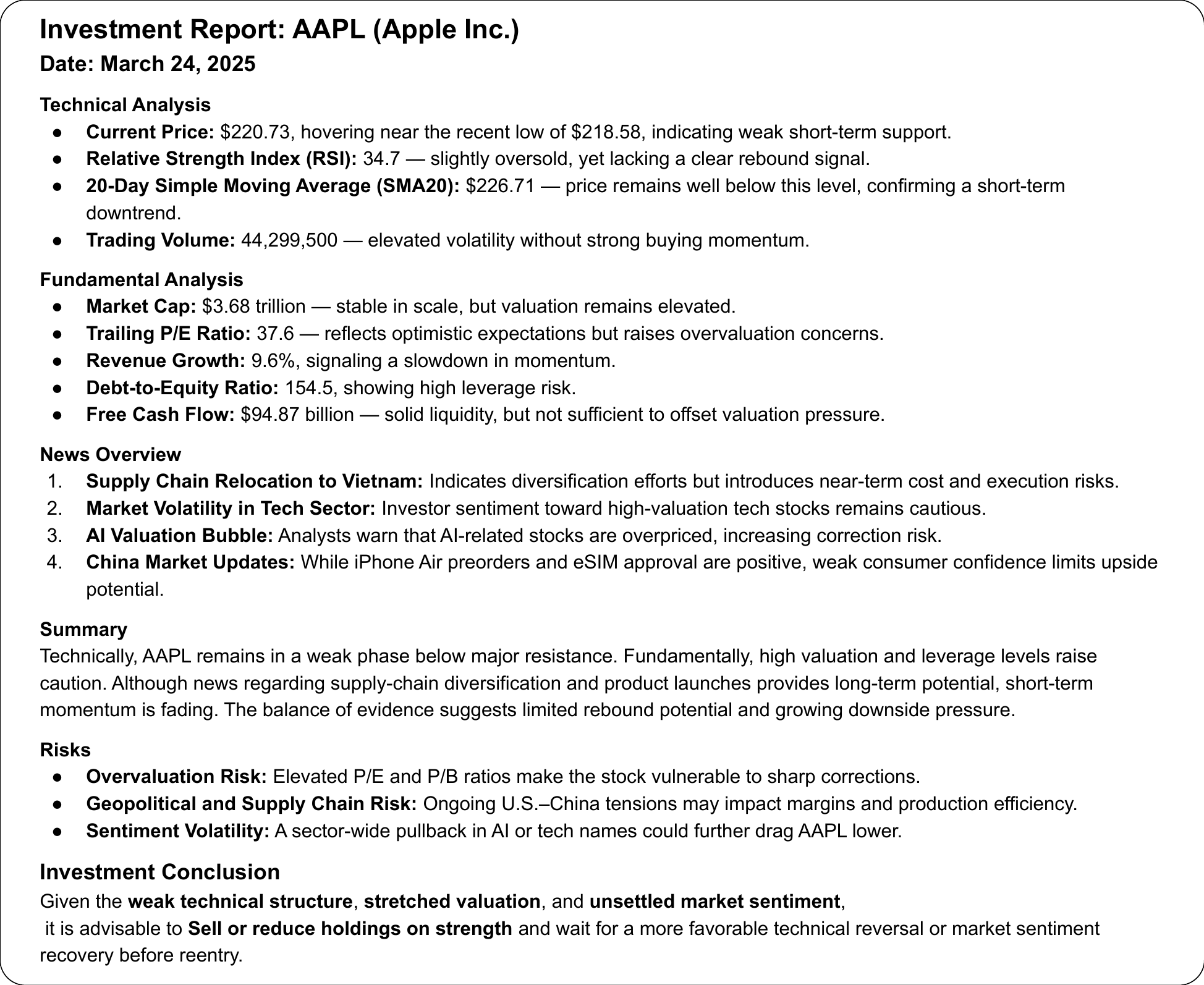}
    \caption{
    Case 1 — \textbf{AAPL Investment Report (March 24, 2025)}.
    P1GPT’s agents collectively issued a \textit{Sell} recommendation as technical weakness, elevated valuation, 
    and negative news sentiment converged. 
    The reasoning emphasized overvaluation risks, geopolitical uncertainty, and lack of short-term momentum.
    }
    \label{fig:aapl_case1}
\end{figure*}

\begin{figure*}[t]
    \centering
    \includegraphics[width=0.95\linewidth]{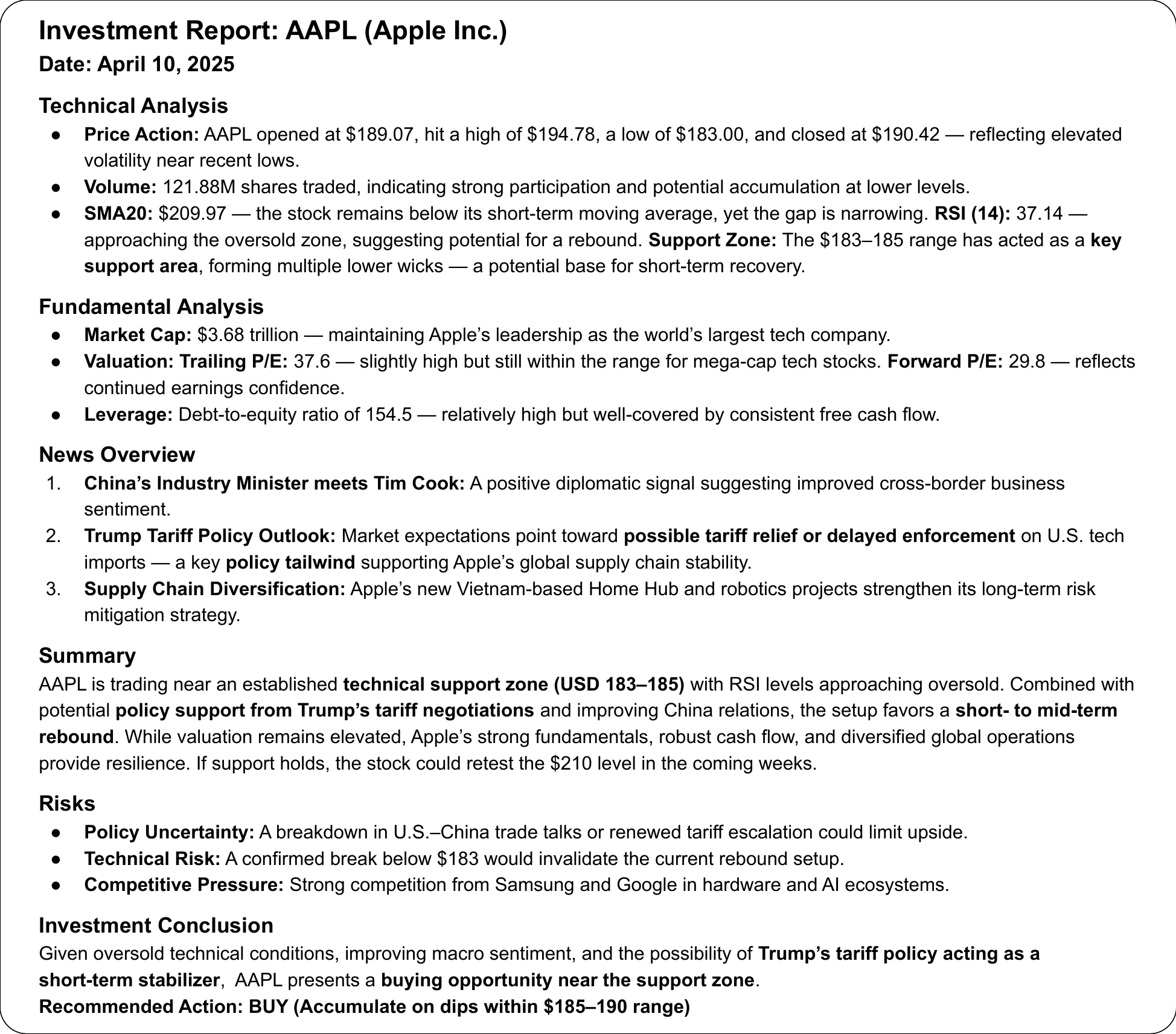}
    \caption{
    Case 2 — \textbf{AAPL Investment Report (April 10, 2025)}.
    Following market stabilization, P1GPT reversed its stance to a \textit{Buy} signal, 
    citing technical rebound near the \$185 support zone, positive policy outlook on U.S.--China tariffs, 
    and improving sentiment indicators. 
    This shift demonstrates the model’s adaptive reasoning and multi-modal evidence integration.
    }
    \label{fig:aapl_case2}
\end{figure*}

To further illustrate P1GPT’s reasoning-driven decision process, we analyze two consecutive investment reports 
generated during March–April~2025 for \textbf{AAPL}.  
These cases demonstrate how the system synthesizes technical, fundamental, and news-based information 
to adapt its stance from a defensive \textit{Sell} position to a confident \textit{Buy} signal as market context evolves.

As shown in Figure~\ref{fig:aapl_case1}, P1GPT initially adopted a cautious outlook on March~24,~2025.  
The \textbf{Technical ISA} detected weak short-term support near \$220, with RSI below 35 and price action 
well beneath the 20-day moving average, signaling persistent downward pressure.  
The \textbf{Fundamental ISA} noted an elevated valuation (trailing P/E~37.6) and high leverage (debt-to-equity~154.5), 
while the \textbf{News ISA} emphasized geopolitical uncertainty, cost risks from supply-chain relocation, 
and broader concerns over a possible AI valuation bubble.  
The consensus among agents concluded that AAPL was in a technically fragile phase with stretched valuation and fading momentum, 
leading to a \textit{Sell or reduce holdings on strength} recommendation.  
This reasoning aligned with the prevailing macro backdrop at the time—renewed tariff tensions and cautious investor sentiment 
toward overvalued technology stocks.

However, as shown in Figure~\ref{fig:aapl_case2}, the system reversed its position on April~10,~2025 
as new information shifted the macro and technical landscape.  
The \textbf{Technical ISA} observed price stabilization near the \$183–185 support zone 
and RSI levels approaching oversold territory (\textasciitilde37), 
indicating the potential for a rebound.  
Simultaneously, the \textbf{News ISA} reported positive developments, including a meeting between China’s Industry Minister and Tim Cook 
and expectations of tariff relief under the Trump administration—signals of improving policy tone and investor sentiment.  
The \textbf{Fundamental ISA} confirmed Apple’s continued profitability (operating margin~29.9\%) 
and robust free cash flow (\$94.9~B), supporting valuation resilience.  
Integrating these updates, the agents collectively upgraded the outlook to \textit{Buy (accumulate on dips)}, 
anticipating a short-term recovery toward the \$210 resistance level.  
This transition from caution to accumulation exemplifies P1GPT’s ability to perform contextual reasoning—reassessing multi-modal inputs dynamically 
rather than following static momentum cues.

These paired cases highlight the explainable nature of P1GPT’s reasoning framework.  
The system’s decisions are not reactive to isolated indicators but emerge from the consensus of specialized agents evaluating 
technical exhaustion, valuation shifts, policy context, and sentiment direction in unison.  
The causal progression from \textit{Sell} to \textit{Buy} demonstrates reasoning coherence and adaptability under uncertainty, 
illustrating how P1GPT’s multi-agent architecture translates real-world evidence into interpretable and strategically consistent actions.

\subsection{Discussion}

Our findings indicate that P1GPT’s edge derives from mechanism design rather than indicator chasing: 
a modular division of labor, standardized inter-agent reporting, and a synthesis step that arbitrates evidence across modalities. 
This design yields an operational principle of \emph{contextual exposure}: the system accepts risk only when multi-modal signals are aligned, 
and scales back when they conflict.
Behaviorally, the agent exhibits stable patterns—measured re-entry after drawdowns and disciplined exits around regime shifts—
that are \emph{traceable} to agent rationales rather than emergent from opaque heuristics. 
Hence, qualitative transparency complements quantitative gains and supports auditability.
Limitations remain. We abstract away transaction costs and portfolio interactions, and evaluate within a single market window. 
Extending to multi-asset allocation, realistic execution frictions, and formal causal tracing from agent outputs to actions 
are necessary steps to assess robustness and scalability.

\section{Conclusion}

We presented \textbf{P1GPT}, a layered multi-agent LLM workflow for multi-modal financial analysis that turns heterogeneous evidence into 
auditable trading decisions. The framework emphasizes modular specialization, structured communication, and integration-time reasoning— 
a combination that delivers interpretable decisions with competitive risk–return characteristics. Future work will focus on deployment realism (cost-aware execution, live paper-trading), portfolio-level extensions (multi-asset, hedging, and dynamic sizing), and deeper transparency (agent-weight adaptation and causal provenance of decisions). We view reasoning-centered, modular architectures as a promising direction for trustworthy financial AI.

\clearpage
\bibliographystyle{unsrt}  
\bibliography{references}  

\end{document}